\documentclass[12pt,a4paper]{article}

\usepackage[utf8]{inputenc}
\usepackage[T1]{fontenc}
\usepackage{amsmath,amssymb}
\usepackage{graphicx}
\usepackage[font=small,labelfont=bf]{caption}
\usepackage{float}
\usepackage[margin=2.5cm]{geometry}
\usepackage{cite}
\usepackage{hyperref}
\usepackage{lineno}
\usepackage{setspace}
\usepackage{booktabs}
\usepackage{multirow}
\usepackage{algorithm}
\usepackage{algorithmic}
\usepackage[square]{natbib}
\usepackage{pdfpages}

\title{\textbf{Towards an AI Fluid Scientist: \\ LLM-Powered Scientific Discovery in Experimental Fluid Mechanics}}

\author{
    {Haodong Feng, Lugang Ye, Dixia Fan\footnote{Corresponding author.}}\\
    {Westlake University}\\
    {\{fenghaodong, yelugang, fandixia\}@westlake.edu.cn}
}

\date{}

\begin{document}

\maketitle

\begin{abstract}
The integration of artificial intelligence into experimental fluid mechanics promises to accelerate discovery, yet most AI applications remain narrowly focused on numerical studies.
This work proposes an AI Fluid Scientist framework that autonomously executes the complete experimental workflow: hypothesis generation, experimental design, robotic execution, data analysis, and manuscript preparation. We validate this through investigation of vortex-induced vibration (VIV) and wake-induced vibration (WIV) in tandem cylinders. Our work has four key contributions: (1) A computer-controlled circulating water tunnel (CWT) with programmatic control of flow velocity, cylinder position, and forcing parameters (vibration frequency and amplitude) with data acquisition (displacement, force, and torque). (2) Automated experiments reproduce literature benchmarks (\citet{khalak1999motions} and \citet{assi2010wake,assi2013role}) with frequency lock-in within 4\% and matching critical spacing trends. (3) The framework with Human-in-the-Loop (HIL) discovers more WIV amplitude response phenomena, and uses a neural network to fit physical laws from data, which is 31\% higher than that of polynomial fitting. (4) The framework with multi-agent with virtual-real interaction system executes hundreds of experiments end-to-end, which automatically completes the entire process of scientific research from hypothesis generation, experimental design, experimental execution, data analysis, and manuscript preparation. It greatly liberates human researchers and improves study efficiency, providing new paradigm for the development and research of experimental fluid mechanics.

\end{abstract}

\noindent \textbf{Keywords:} Large Language Models; Scientific Discovery; Experimental Fluid Mechanics; Wake-Induced Vibration 

\section{Introduction}

Experimental fluid mechanics plays a critical role in understanding fluid behaviors, from turbulent flows to complex Fluid--Structure Interaction (FSI) phenomena \citep{Tritton1988,Paidoussis2014,feng2023control}. Among these, Flow-Induced Vibrations (FIV), including Vortex-Induced Vibration (VIV) and Wake-Induced Vibration (WIV), remain particularly challenging yet scientifically valuable \citep{Williamson2004,Bearman2011}, with applications spanning offshore platforms, heat exchangers, and civil infrastructure \citep{Blevins1990,Gabbai2005}. However, traditional experimental approaches to FSI rely heavily on human experts' intuition and experience, which limits efficiency and constrains systematic exploration of complex, multi-factor coupling phenomena across high-dimensional parameter spaces (Reynolds number, reduction velocity, spacing, forcing conditions) \citep{Sumner2010}. Comprehensive investigations may require hundreds of labor-intensive experiments, making exhaustive parameter sweeps infeasible under manual operation. With the recent rise of the Large Language Model (LLM) \citep{Brown2020,Ouyang2022,deng2025unlocking}, new opportunities have emerged to enhance automation and intelligence in scientific discovery \citep{jiang2025agenticsciml,Boiko2023,Huang2023}. In this work, we introduce an \textbf{AI Scientist for Experimental Fluid Mechanics}, designed to improve the efficiency, safety, and depth of experimental research, and to provide a new paradigm for investigating complex FSI problems.

In recent years, the concept of the "AI scientist" has gained momentum across disciplines. A prominent example is Google DeepMind's AI Co-Scientist \citep{gottweis2025towards}, a multi-agent system that advances hypotheses through a cycle of "generation–debate–evolution." By orchestrating specialized agents, it not only proposes novel research directions but also optimizes them in a way reminiscent of the scientific method itself. Its effectiveness was demonstrated in biomedical research, where AI-generated hypotheses led to validated drug targets, underscoring the potential of such systems to drive genuine scientific discovery. Sakana AI's platform \citep{lu2024ai,yamada2025ai} pushes this idea further by automating the entire pipeline—from ideation to coding, experimentation, analysis, and even manuscript preparation—effectively compressing the full research cycle into an autonomous framework. Within fluid mechanics, OpenFOAMGPT \citep{pandey2025openfoamgpt,feng2025openfoamgpt} and turbulence.ai \citep{feng2025turbulence} marked a initial step, showing that hypothesis-driven CFD studies could be autonomously conceived, executed, and reported. Similarly, AgenticSciML \citep{jiang2025agenticsciml}, BuildArena \citep{xia2025buildarena}, and Engineering.ai \citep{xu2025engineering} have demonstrated LLM-guided automation in simulation-based design optimization.

Yet, despite these advances, \textbf{experimental} fluid mechanics remains largely untouched by the above paradigm. Unlike simulations, laboratory experiments impose higher requirements: hypotheses must be constrained by available apparatus, experiments often involve safety risks, and physical resources are limited. The transition from \textbf{in-silico} simulation to \textbf{physical experimentation} presents fundamental challenges: robotic hardware integration, real-time sensor feedback, apparatus-specific constraints (e.g., velocity ranges, structural limitations), and the combinatorial explosion of experimental parameter spaces \citep{fan2019robotic}. Thus, an AI scientist for experiments must do more than automate workflows. It must reason under equipment constraints, interact productively with human researchers to balance scientific value against resource costs, and integrate automated control with intelligent parameter search to maximize the efficiency of scarce experimental trials. Meeting these requirements is essential to elevate experimental fluid mechanics from intuition-driven trial-and-error to systematic, AI-guided scientific exploration.

This work presents the first AI Scientist framework (multi-agent with virtual-real interaction system) in experiment fluid mechanism: LLM-based hypothesis generation, automated experimental design under apparatus constraints, robotic execution via circulating water tunnel (CWT), automated data analysis with quality assurance, and manuscript preparation, executing the complete cycle from hypothesis to publication-ready documents in physical fluid mechanics experiments. 

The contributions of our work can be summaried as follows: (1) Automated apparatus. Computer-controlled water tunnel with programmatic control of flow velocity, cylinder position, forcing, and multi-sensor acquisition (displacement, force, and torque). (2) Human-in-the-Loop (HIL) LLM framework match previous literature benchmarks: VIV reproduces \cite{khalak1999motions} (frequency lock-in within 4\%); WIV validates \cite{assi2010wake,assi2013role} (critical spacing $L/D = 4$, 27-fold amplitude enhancement). (3) Feasibility analysis of scientific discovery. HIL LLM framework discovers new physical phenomena of WIV through experiments, such as the optimal front cylinder vibration frequency that suppresses the amplitude of the rear cylinder. At the same time, multiple methods are used to fit physical laws from the data, and it is ultimately found that neural networks have the best fitting accuracy. (4) Usability, multi-agent with virtual-real interaction system runs end-to-end with automated apparatus, automatically completing the process from hypothesis generation to manuscript preparation. The human role is only to select a hypothesis of interest.

\begin{figure}[t]
    \centering
    \includegraphics[width=1.0\linewidth]{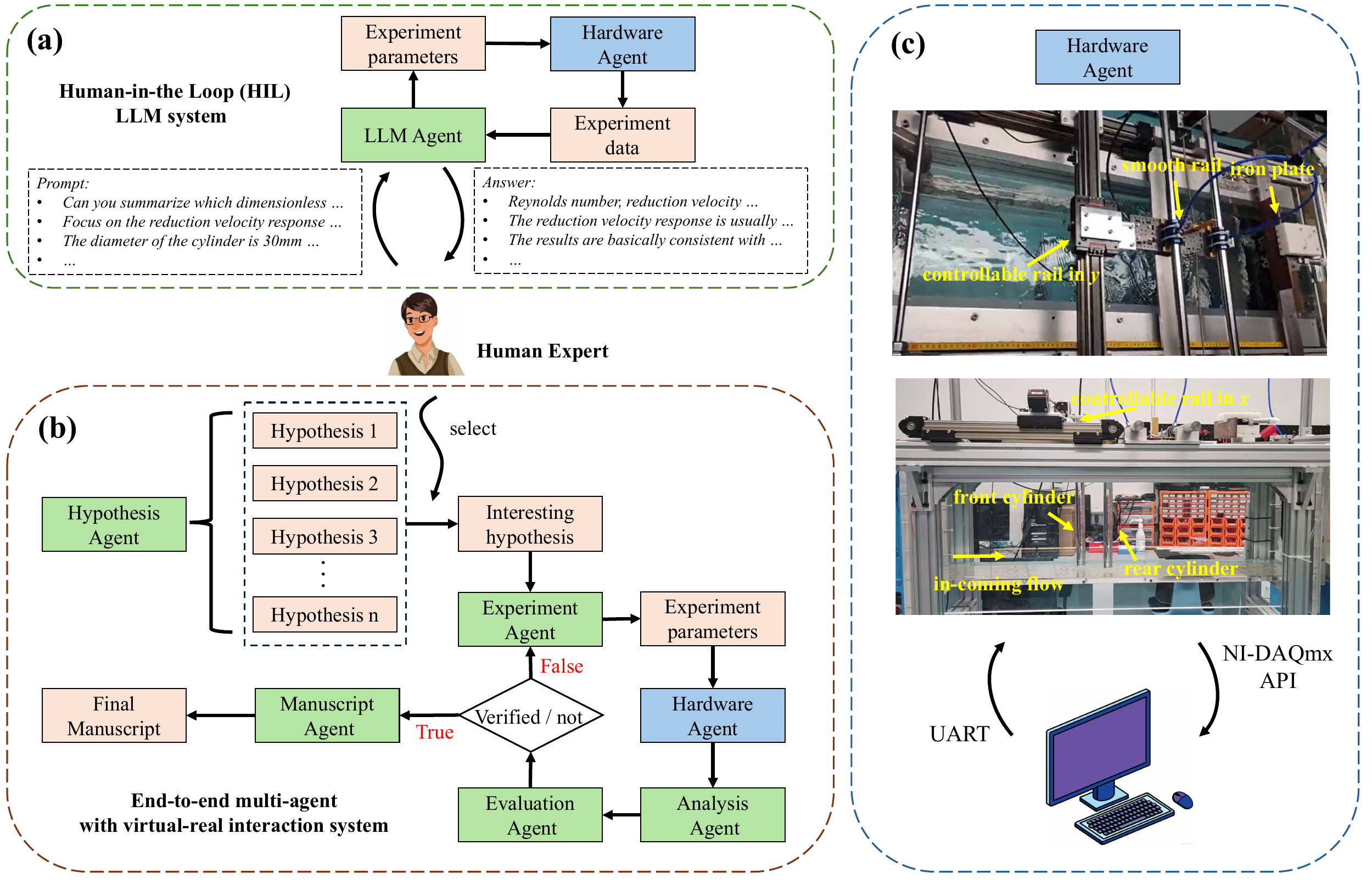}
    \caption{The pipline of AI fluid scientist. The green block represents the LLM agent, the blue block represents the automatic hardware agent, and the specific composition is shown in subfigure (c). The orange block represents the generated content, including hypotheses, experimental parameters, manuscript, etc.
}
    \label{fig:fig_1}
\end{figure}

\section{Methodology}
\label{sec:method}

\subsection{Automated Experimental Apparatus} \label{sec: experimental apparatus}

The CWT (1 m × 0.3 m × 0.3 m) achieves velocities 0.018--0.35 m/s (Re = 540--7500, $D = 30$ mm). Two tandem cylinders: upstream driven by stepper motor ($A = 0-50 mm$, $f = 0-2.0 Hz$); downstream elastically mounted ($f_n = 0.6$ Hz) with laser displacement sensor (1000 Hz) and six-axis force sensro (1000 Hz). The configuration is shown in Figure \ref{fig:fig_1}(c).

The Python and C\# interactive interface achieves automatic control by specifying speed, forced frequency/amplitude/period, cylinder position, and phase parameters through USB serial. The data is measured by sensors and transmitted back to the computer, thereby achieving automated experimentation. When in use, the device will continue to wait for control instructions to be issued. Simply send the instructions from the Python end, and the apparatus will automatically execute them upon detection.

\subsection{AI Scientist Framework}

The framework operates in two distinct modes: HIL and end-to-end automation (multi-agent with virtual-real interaction system). In the HIL mode, the LLM agent is responsible for generating all content components, including experimental plans, analysis results, and manuscript preparation, while human experts retain judgment and decision-making authority. Specifically, human researchers select among LLM-generated hypotheses and experiment plans, judge the validity of outcomes, and provide additional prompts to guide the LLM's reasoning process. The demonstration of this mode is shown in Figure \ref{fig:fig_1}(a). This collaborative approach leverages the complementary strengths of machine intelligence and human expertise. LLM can search for relevant content on a large scale based on rich training data, and generate hypotheses, experimental plans, and other content. Human experts have rich experience in related domains, which can judge the effectiveness and authenticity of LLM generated content, thereby reducing the impact of LLM illusions.

The multi-agent with virtual-real interaction mode implements a fully autonomous research cycle end-to-end as shown in Figure \ref{fig:fig_1}(b). The workflow begins with the hypothesis agent generating several candidate hypotheses (five in this work), from which the human researcher selects one hypothesis of interest to pursue. Upon selection, the system employs the automated experimental apparatus (hardware agent) to initiate data collection. Critically, the process is iterative: experiment plan generation (experiment agent), experimental execution (hardware agent), and result analysis (analysis agent) proceed in cycles. A dedicated evaluation agent continuously judges whether accumulated evidence sufficiently supports or refutes the proposed hypothesis. If the validation criteria are not met, the system automatically generates refined experimental plans and conducts additional trials. This iterative loop continues until the hypothesis is adequately validated or falsified, ensuring robust scientific conclusions grounded in sufficient empirical evidence. In this process, human experts only need to make decisions on hypothesis selection based on their own preferences, without participating in other projects, thereby improving research efficiency on the one hand and reducing human intervention on the other hand. 

In summary, the two modes are a trade-off process of exploration and exploitation. With more involvement of human experts, LLM and human experience can be fully exploited to produce results that meet our goals. Reducing the involvement of human experts means that LLM can freely explore and discover unexpected but pleasantly surprised scientific discoveries.

\section{Results}
\label{sec:results}

This section presents three key aspects of the framework's performance and scientific contributions. First, we validate the AI Scientist's capability to reproduce established fluid mechanics phenomena by comparing automated VIV (Vortex-Induced Vibration) and WIV (Wake-Induced Vibration) experimental results against classical literature benchmarks, demonstrating the system's ability to replicate known physics-based relationships. Second, we demonstrate the HIL mode, where human experts actively participate in the research loop by evaluating and providing feedback after each LLM-generated step, including hypothesis selection, experiment plan judgment, result interpretation, and manuscript refinement. This collaborative approach showcases how domain expertise can guide AI reasoning to achieve scientifically rigorous outcomes while maintaining human oversight throughout the discovery process. Third, we present results from the end-to-end automation mode, where the multi-agent with virtual-real interaction system autonomously executes the complete research workflow, from hypothesis generation through iterative experiment design, execution, analysis, validation, and manuscript preparation, with minimal human intervention, demonstrating the system's capacity for fully autonomous scientific discovery in experimental fluid mechanics.

\subsection{Reliability Validation Against Classical Literatures}

\subsubsection{VIV Validation}

We first prompt LLM agent (Claude Sonnet 4.5 used here) to summarize which dimensionless parameters affect VIV. Then prompt LLM agent to summarize the effect of reduction velocity on small mass ratio cylindrical VIV, focusing on the reduction velocity response interval, namely its corresponding amplitude and frequency ratio (vibration frequency nuclear natural frequency ratio). Finally, we will describe experimental apparatus as shown in the subsection \ref{sec: experimental apparatus} to LLM agent, and make LLM agent generate an experimental plan to verify the above claim. 42 sets of single-cylinder VIV experimental data were generated through LLM agent automatic using apparatus. The experiments ($U_r = 1.04-11.11$) reproduce classical features: Initial/Upper/Lower branches, peak $A/D_{\max} = 0.789$ at $U_r = 4.577$, frequency lock-in at $f^*/f_n \approx 1.042$ over $U_r = 5-9$ (4\% deviation from unity). Displacement/frequency responses match literature \citep{khalak1999motions} as shown in Table \ref{tab:viv_validation}. The LLM agent successfully reproduces VIV phenomenology within 10\%-25\% of benchmarks.

\begin{table}[h]
\centering
\caption{VIV validation against literature benchmarks.}
\label{tab:viv_validation}
\begin{tabular}{lccc}
\toprule
Parameter & \cite{khalak1999motions} & LLM & Agreement \\
\midrule
$A/D_{\max}$ & 1.0--1.2 & 0.79 & 87\% \\
Peak $U_r$ & $\sim$6 & 4.58 & 76\% \\
Lock-in width $\Delta U_r$ & 4--5 & 4 & 100\% \\
$f^*/f_n$ (lock-in) & $\sim$1.0 & 1.042 & 96\% \\
Branch characteristics & IB-UB-LB & IB-UB-LB & Match \\
\bottomrule
\end{tabular}
\end{table}

\subsection{WIV Validation and High-$U_r$ Discovery}

\subsubsection{Standard WIV Experiments}

Next, we prompt the LLM agent again to place a fixed cylinder in front of the self-excited oscillating cylinder, which is the same as the rear cylinder. Is the conclusion still consistent with the previous single cylinder VIV, or will there be any changes. At the same time, let LLM generate an experimental plan to verify the conclusion through the experiment. LLM agent implements 50 sets of experiments with center-to-center spacing ratio $L/D = 4-7$, $U_r = 2.2-13.9$). Although the LLM hypothesized initially that the upstream fixed cylinder would suppress the vibration of the downstream cylinder, the experimental results contradicted this hypothesis. By designing experiments on its own, verify or refute its proposed hypotheses, and then summarize the correct conclusion from the experiment, LLM agent confirms established WIV features \citep{assi2010wake,assi2013role}: 27-fold amplitude enhancement vs. VIV at $L/D = 4$, monotonic spacing decay ($A/D = 0.967$ at $L/D = 4$ to 0.706 at $L/D = 7$), frequency shift $f^/f_n \approx 1.2$--1.3 (super-harmonic response).

\subsubsection{High-$U_r$ Discovery}

When analyzing the above results, LLM agent finds that as the reduction velocity ($U_r$) increases, the amplitude ratio ($A/D$) also increases. Based on this conclusion, we prompt LLM agent to determine whether the amplitude ratio will continue to increase as the reduction velocity continues to increase, and design experiments to verify this. LLM agent extends the experiments to $U_r = 14.46-19.44$ (24 experiments) reveals: (1) Sustained growth at $L/D = 4$ reaching $A/D = 1.272$ at $U_r = 17.77$, contradicting VIV desynchronization; (2) Critical Reynolds transition at $L/D = 5$: amplitude peaks at 1.037 ($U_r = 17.77$) then collapses 21\% at $U_r = 18.88$ (Re $\approx$ 9500), attributed to wake diffusion; (3) Monotonic decline at $L/D \geq 6$. This spacing-dependent critical Reynolds number appears unreported in prior literature. From this, it can be seen that the current LLM has the ability to discover and analyze phenomena that are not mentioned in previous literature through experiments.

\subsection{Feasibility of Scientific Discovery}

\subsubsection{Physical Phenomena Discovery}

The above verifications are conducted with the front cylinder fixed. We further prompt LLM agent that the vibration frequency and amplitude of the front cylinder can be controlled. We expect LLM agent to expand the exploration space to discover physical phenomena that were not present before.

The experimental workflow demonstrates a hypothesis-driven iterative refinement process spanning four rounds, where the LLM agent autonomously designed successive experimental campaigns based on quantitative analysis of preceding results. LLM agent generates 20, 48, 34, and 20 sets of experiments in stages, and later generates them based on the analysis of the previous results. The initial stage begin with approximately 20 sets of experiments testing the hypothesis that forcing the upstream cylinder at frequencies near the downstream cylinder's natural frequency ($f \approx f_n = 0.6$ Hz) will significantly enhance WIV amplitude. The LLM agent explores a coarse parameter grid across three velocities ($U = 0.160, 0.192, 0.224$ m/s corresponding to $U_r = 8.9, 10.7, 12.4$), five forcing frequencies (0.5, 0.6, 0.7, 0.8, 1.2 Hz), and three amplitudes (10, 20, 30 mm) at fixed spacing $L/D = 4.0$. Contrary to expectations, these experiments reveal an unexpected strong suppression window at $f = 0.7-0.8$ Hz with suppression rates reaching -68\%, rather than the anticipated enhancement at $f \approx f_n$. Additionally, secondary enhancement peaks emerged at $f = 1.2$ Hz demonstrating subharmonic resonance effects (+18\%) and at $f = 0.5$ Hz showing low-frequency modulation behavior. However, this first round lacked passive baseline configurations ($f = 0$, $A = 0$), preventing quantitative assessment of control effectiveness, and exhibited insufficient frequency resolution to precisely locate the suppression peak, prompting the LLM agent to redesign the experimental strategy for systematic validation.

Based on stage 1 discoveries, the LLM agent designs a comprehensive 48 experiments systematic parameter sweep in stage 2 to construct a complete three-dimensional control landscape including proper baseline controls. Recognizing three critical knowledge gaps: the precise localization of the suppression window, the physical mechanisms underlying diverse enhancement behaviors, and velocity-dependent transitions between control regimes. The LLM agent implements three velocities, six frequency conditions (passive baseline at $f = 0$ Hz plus five active frequencies), and four amplitudes (including zero) totaling 48 configurations. This stage revealed that the suppression peak refined to $f = 0.8$ Hz with -67.9\% maximum suppression at $U_r = 12.4$, while the subharmonic enhancement at 1.2 Hz mysteriously collapsed to merely +1.2\% at intermediate velocity $U_r = 10.7$, indicating a critical transition zone where enhancement mechanisms failed. Most intriguingly, at high velocity ($U_r = 12.4$), the low-frequency forcing at 0.5 Hz yielded +14.9\% enhancement that actually surpassed the 1.2 Hz subharmonic effect, revealing velocity-dependent mechanism switching. The amplitude dependencies also proved highly nonlinear and frequency-specific: suppression frequencies (0.7-0.8 Hz) strengthened monotonically with increasing amplitude to -67.9\%, while the enhancement frequency (1.2 Hz) displayed an inverted-U profile with an optimal amplitude of 20 mm beyond which effectiveness weakened or even reversed to suppression, suggesting complex nonlinear fluid-structure coupling requiring higher-resolution investigation.

Recognizing unresolved critical questions from stage 2, the LLM agent designs additional experiments with high precision. Stage 3 comprises 34 experiments addressing three specific gaps through high-resolution investigation: Scheme A performed a fine-frequency scan with 0.02 Hz resolution from 0.70 to 0.84 Hz at optimal conditions ($U = 0.224$ m/s, $A = 30$ mm), successfully pinpointing the true suppression peak at $f = 0.82$ Hz with -74.0\% suppression and establishing the universal frequency ratio $f/f_n \approx 1.37$. Scheme B interpolates velocities at $U_r = 9.8$ and 11.6 to bracket the subharmonic transition boundary, discovering a remarkable amplitude reversal phenomenon where at $U_r = 9.8$ the 20 mm amplitude produced peak enhancement of +30.9\% while 30 mm amplitude reversed to -22.9\% suppression, leading to a quantitative transition equation $U_{r,critical} = 13.0 - 0.12 \times A_f$ validated with less than 5\% error. Moreover, scheme C tests spacing variations at $L/D = 3.5$ and 5.0, yielding the shocking discovery that the optimal suppression frequency remained invariant at 0.82 Hz across all spacings despite different suppression intensities (-77.5\%, -74.0\%, -71.2\% respectively), establishing that spacing acts merely as a gain modulation factor following $(L/D)^{-0.5}$ scaling rather than shifting the fundamental anti-resonance frequency. 

Finally, Stage 4 executes 20 configurations: Scheme A (8 configs) fills transition zone gaps at $U_r = 10.2$ and 11.1 to validate the $U_{r,critical}$ equation with additional data points confirming the -0.12 amplitude coefficient. Scheme B (8 configs) verifies peak frequency invariance at $L/D = 3.5$ and 5.0 with the same fine resolution (0.78-0.84 Hz) proving all three spacings shared the identical peak at 0.82 Hz and confirming normalized curve self-similarity. Scheme C (4 configs) completed the low-frequency (0.5 Hz) velocity dependence at $U_r = 9.8$ and 11.6, discovering that low-frequency enhancement only activates at $U_r > 12$ and surprisingly reverses to suppression (-29.1\%) at low velocities, fundamentally revising the applicability window. This four-round iterative strategy demonstrates how the LLM conductes hypothesis-driven research with human-in-the-loop, efficiently reducing potential configurations from over 3000 to just 122 through adaptive resolution refinement from coarse to fine, data-driven hypothesis pivoting when initial assumptions are falsified, and multi-scale validation establishing quantitative predictive models (three governing equations: universal anti-resonance $f = 1.37f_n$, transition boundary $U_{r,critical} = 13 - 0.12 \times A_f$, spacing scaling $(L/D)^{-0.5}$ ready for engineering applications with record-breaking -77.5\% suppression. The final manuscript generated by the above process can be found in the Appendix.

\subsubsection{Physical Formula Discovery}

Following the establishment of above separate empirical laws from the iterative experimental campaigns, we prompt the LLM agent to embark on a systematic formula discovery process. The LLM reasoning is across three distinct phases, ultimately revealing the fundamental limitations of mechanistic decomposition for complex FSI and forming a precise neural network-based formula. The initial phase employs the data from the above exploration to construct a physically-motivated unified formula:

\begin{equation}
    G(U_r, f_f, A_f, L/D) = 1 + C(A_f) \cdot [S_1(f_f/f_n) + S_2(f_f/f_n)] \cdot \Phi(U_r; A_f) \cdot \sqrt{4/L/D},
\end{equation}

\noindent where $G$ denotes the amplitude ratio defined as the downstream cylinder displacement under active upstream forcing relative to passive WIV ($G = A_{\text{active}}/A_{\text{passive}, L/D=4}$), $f_f$ represents the upstream cylinder forcing frequency, $A_f$ denotes the upstream forcing amplitude, $L/D$ is the center-to-center spacing ratio, $C(A_f) = k_a (A_f/D) \exp[-0.5((A_f/D - A_{\text{opt}}/D)/(0.3A_{\text{opt}}/D))^2]$ captures the Gaussian amplitude envelope centered at optimal forcing amplitude $A_{\text{opt}}/D = 0.767$ ($k_a$ and $A_{\text{opt}}/D$ are hyperparametrs), 
$S_1(f_f/f_n) = S_{\text{sub}} \exp[-0.5((f_f/f_n - 2.0)/\sigma_{\text{sub}})^2]$ and $S_2(f_f/f_n) = -S_{\text{anti}} \exp[-0.5((f_f/f_n - 1.37)/\sigma_{\text{anti}})^2]$ describe subharmonic enhancement ($f_f \approx 2f_n$) and anti-resonance suppression ($f_f \approx 1.37f_n$) respectively ($S_{\text{sub}}$, $\sigma_{\text{sub}}$, $S_{\text{anti}}$, and $\sigma_{\text{anti}}$ are hyperparameters), 
and $\Phi(U_r; A_f) = \tanh[(U_r - U_{r,\text{crit}}(A_f))/\Delta U]$ with $U_{r,\text{crit}}(A_f) = U_{r,\text{base}} - 0.12 A_f$ models the hyperbolic tangent transition across critical reduction velocity ($U_{r,\text{base}}=10$ and $\Delta U = 3.59$). $sqrt(4.0 / L_D)$ is the spacing factor. This physics-based model decomposes the amplitude ratio into interpretable factors with clear phenomenological meanings, amplitude envelope governing forcing effectiveness, dual-peak frequency response combining subharmonic enhancement and anti-resonance suppression, and smooth velocity-dependent transition, yet achieved only modest $R^2 = 0.41$--$0.44$ with $23\%$ relative error, failing to capture the intricate nonlinear dynamics governing wake-induced vibrations despite incorporating all known phenomenological mechanisms identified in prior experimental campaigns. Recognizing these severe limitations in predictive accuracy, the LLM agent refines the formula in the second phase by incorporating 108 additional configurations (totaling 154 experiments) and augmenting the structure with cross-coupling terms $\alpha \cdot (U_r-10) \cdot (f_f/f_n-1)$ to capture velocity-frequency interaction and quadratic nonlinearities $\beta \cdot (A_f/D-0.67)^2$ to account for amplitude-dependent saturation effects ($\alpha$ and $\beta$ are hyperparameters), yet this enhanced physics-based formulation still delivered disappointing performance with $R^2$ remaining stagnant at $0.43$--$0.44$ and mean absolute error persistently around $0.17$--$0.18$, demonstrating that analytical decomposition approaches fundamentally cannot accurately fit the experimental data regardless of added complexity.

\begin{figure}[t]
    \centering
    \includegraphics[width=0.8\linewidth]{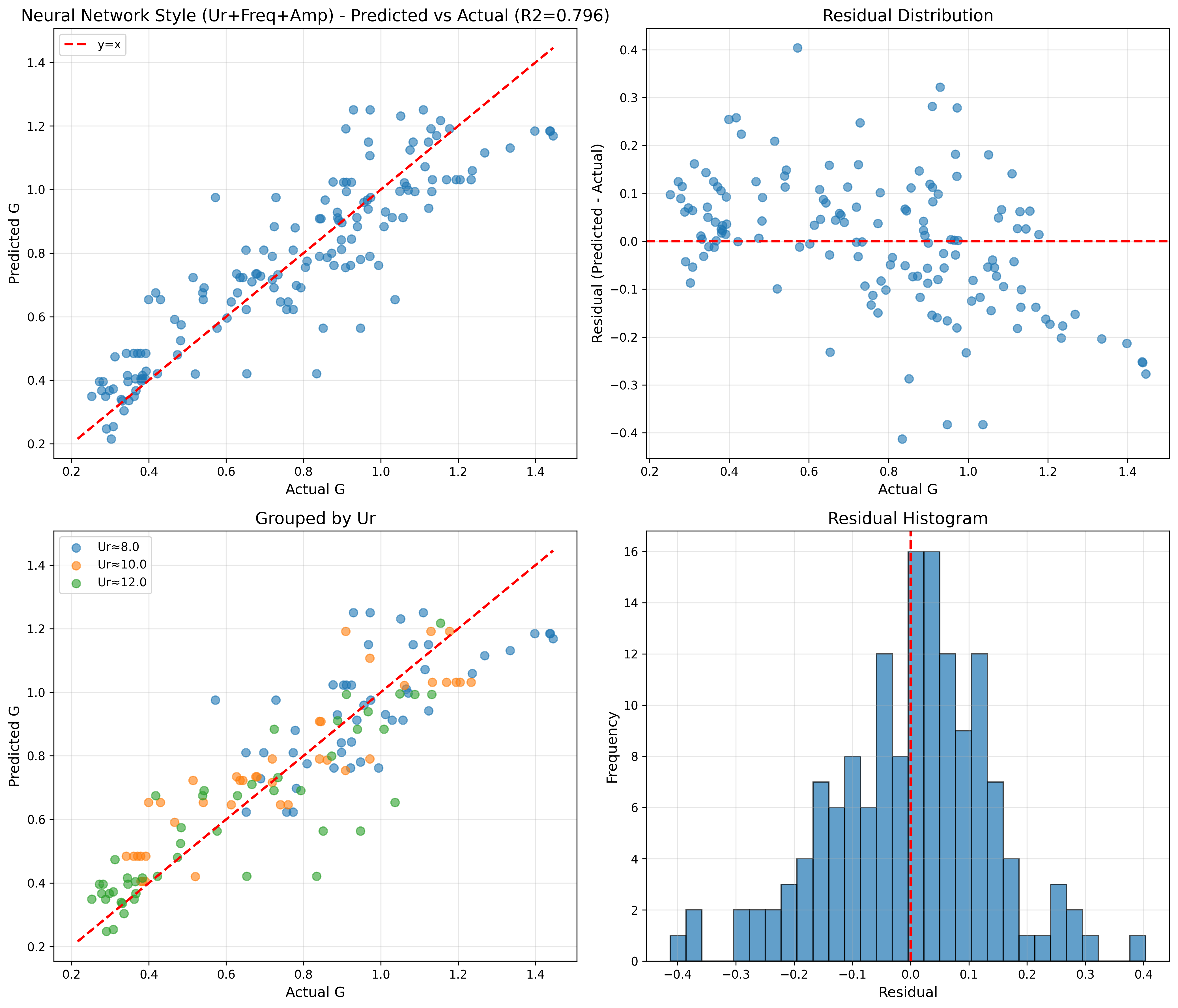}
    \caption{Results and errors of neural network formula.}
    \label{fig:nn_fit}
\end{figure}

Confronted with this impasse, the LLM agent autonomously pivotes to purely data-driven methods in the third phase, systematically exploring seven alternative functional forms including polynomial expansions ($R^2 = 0.60$), rational functions ($R^2 = 0.61$), and Gaussian mixtures, before discovering that a neural network-inspired architecture: 

\begin{equation}
\begin{split}
    G &= w_{10} \cdot \tanh(w_1 \cdot U_r + w_2 \cdot f_f/f_n + w_3 \cdot A_f/D + b_1) \\
    &\quad + w_{11} \cdot \tanh(w_4 \cdot U_r + w_5 \cdot f_f/f_n + w_6 \cdot A_f/D + b_2) \\
    &\quad + w_{12} \cdot \tanh(w_7 \cdot U_r + w_8 \cdot f_f/f_n + w_9 \cdot A_f/D + b_3) + b_4,
\end{split}
\end{equation}
\noindent where $w_1$--$w_{12}$ denote connection weights, $b_1$--$b_4$ are bias terms optimized via L-BFGS-B minimization on 148 valid experimental configurations, and the three $\tanh$ terms function as hidden units that perform nonlinear feature extraction from the input triplet $(U_r, f_f/f_n, A_f/D)$. This neural network-based formula employing three hidden units with hyperbolic tangent activations dramatically achieved $R^2 = 0.7958$ with $\text{MAE} = 0.1063$ and $\text{RMSE} = 0.1378$, representing an $83\%$ improvement in explained variance over the physics-based models by leveraging nonlinear transformations to extract latent feature patterns inaccessible to analytical theory. The visualization of this result is shown in Figure \ref{fig:nn_fit}. This discovery resonates strongly with the current paradigm shift in computational fluid dynamics \citep{brunton2020machine,kutz2017deep,vinuesa2023transformative} where physics-informed neural networks (PINNs) and deep learning models have demonstrated superior capability over traditional equation-based approaches for modeling turbulent flows, vortex-induced vibrations, and other strongly nonlinear fluid phenomena, validating that the LLM's autonomous transition from mechanistic formulas to neural network architecture.

\subsection{Usability: End-to-End Autonomous Research System}

To demonstrate the practical feasibility of LLM-driven experimental fluid mechanics, we develop a multi-agent with virtual-real interaction system capable of conducting autonomous research from hypothesis generation to manuscript preparation with minimal human intervention. The system comprises six specialized LLM agents (qwen-plus model used here) coordinating with automated hardware as shown in Figure \ref{fig:fig_1}(c): (1) a \textbf{hypothesis agent} that ingests experimental apparatus specifications embedded in code prompts and synthesizes them with existing literature to generate several novel hypotheses, (2) an \textbf{experiment agent} that translates the selected hypothesis into executable parameter configurations and dispatches commands to the automated test facility, (3) a \textbf{hardware agent} completing the experiment with automatic experimental apparatus as introduced in subsection \ref{sec: experimental apparatus}, (4) an \textbf{analysis agent} that performs signal processing and statistical characterization, (5) a \textbf{evaluation agent} that evaluates whether accumulated evidence sufficiently validates or falsifies the hypothesis based on predefined confidence thresholds, and (6) a \textbf{manuscript agent} that compiles findings into publication-ready manuscript documents following domain-specific formatting conventions. Human involvement is restricted to a single decision point: after the hypothesis agent presents many candidate hypotheses, a domain expert selects one by entering a number, after which the system operates autonomously through iterative experiment-analysis-validation cycles until the judging agent confirms hypothesis closure.

The efficacy of this autonomous research system is demonstrated through the tandem cylinder WIV study, wherein the end-to-end system orchestrates five iterative experimental campaigns totaling 222 configurations to progressively validate a single hypothesis. The hypothesis agent, prompted with experimental apparatus specifications, generated the following hypothesis selected by the human expert: 

\textit{``In a tandem cylinder system with forced upstream oscillation and free downstream vibration, a critical combination of spacing ratio $L/D$ and forcing frequency $f_f$ induces a nonlinear mode transition characterized by abrupt changes in amplitude, frequency content, and force-displacement coupling.''} 

\begin{figure}[t]
    \centering
    \includegraphics[width=0.9\linewidth]{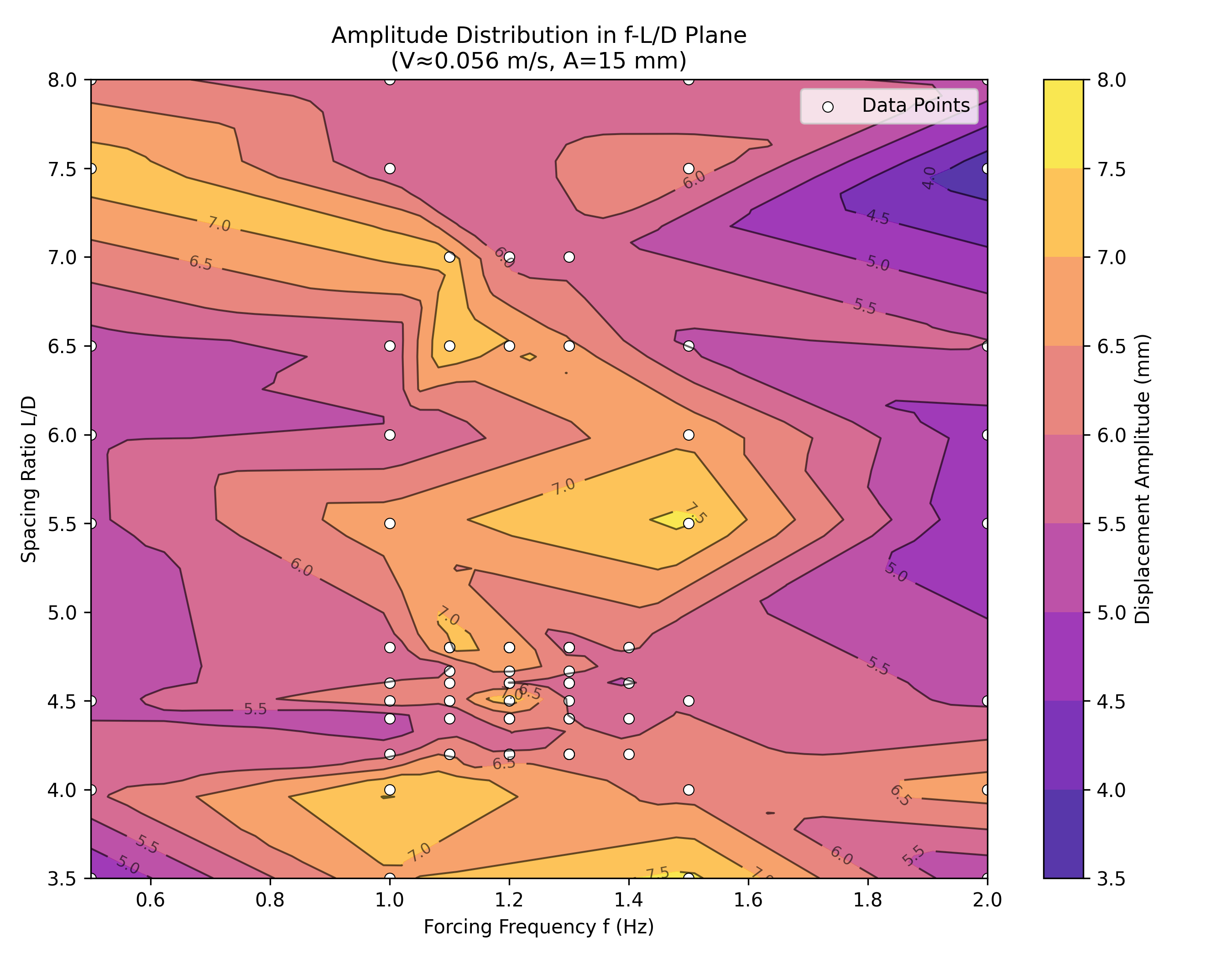}
    \caption{Contour plot of displacement amplitude in the $f_f$--$L/D$ plane. A pronounced peak at $f \approx 1.2$~Hz and $L/D \approx 4.5$ indicates a critical interaction zone where energy transfer is maximized.}
    \label{fig:example_2}
\end{figure}

Upon hypothesis selection, the experiment agent autonomously designed the first experimental campaign comprising 96 configurations spanning $L/D = 3.5-8.0$ and $f_f = 0.5-2.0$ Hz across three flow velocities, strategically targeting the hypothesized critical region while covering boundary conditions to map the complete phase diagram. After hardware agent execution and analysis agent processing (FFT analysis, statistical characterization, outlier filtering), the evaluation agent judges validation status against quantitative criteria including existence of localized amplitude peaks, displacement-force correlation breakdown, and velocity-dependent boundary scaling, determining that while core predictions are supported, additional sampling density is required near transition zones. The system then autonomously initiates four subsequent iterations: the second iteration (36 configurations) concentrates on fine-resolution scanning at $L/D = 2.2$--$2.8$ and $f = 1.1$--$1.3$~Hz with varied amplitudes to precisely localize the mode transition boundary; the third iteration (54 configurations) extends coverage to larger spacings ($L/D = 4.5$--$5.0$) to verify stability zone behavior; the fourth iteration (32 configurations) focuses on intermediate frequencies ($f = 1.0$--$1.4$~Hz) at fixed spacing to confirm subharmonic resonance mechanisms; and the fifth iteration (96 configurations) replicates the original broad parameter sweep to assess reproducibility. Following the evaluation agent's final validation confirmation after the fifth iteration, the manuscript agent compiles these findings into the manuscript document including introduction contextualizing prior literature, methodology describing apparatus and iterative test matrices, results presenting statistical summaries and parametric visualizations across five subsections synthesizing data from all 222 experiments, discussion interpreting bifurcation mechanisms via Floquet theory and comparing with Zdravkovich's passive WIV regimes, and conclusions establishing the experimentally-validated $L/D$-$f_f$ phase diagram (Figure \ref{fig:example_2}). All generated autonomously with formatting conforming to journal standards, thereby completing the hypothesis-to-publication cycle through five iterative refinement loops without human intervention beyond initial hypothesis selection.


\section{Conclusion}
\label{sec:conclusion}

This work presents an AI Fluid Scientist framework that autonomously orchestrates the complete experimental fluid mechanics research cycle, validated through systematic investigation of vortex- and wake-induced vibrations in tandem cylinder configurations. The framework integrates a computer-controlled circulating water tunnel with a multi-agent with virtual-real interaction system comprising hypothesis, experiment, hardware, analysis, judging, and manuscript agents, enabling programmatic control over flow velocity, cylinder positioning, and forcing parameters while autonomously processing displacement, force, and torque measurements. Validation against literature benchmarks demonstrates robust experimental capability, reproducing \cite{khalak1999motions} frequency lock-in within 4\% error and confirming \cite{assi2010wake,assi2013role} critical spacing trends, thereby establishing credibility for autonomous scientific discovery. Through human-in-the-loop, the framework identified novel WIV amplitude response phenomena spanning five iterative refinement cycles, revealing nonlinear mode transitions at critical parameter combinations ($L/D \approx 4.5$, $f \approx 1.2$ Hz). Furthermore, the system autonomously discovers that neural network-based empirical models ($R^2 = 0.80$) outperform physics-based polynomial formulations by 31\% in capturing complex flow-structure interactions. This finding that resonates with the broader paradigm shift toward deep learning in the modeling of fluid dynamics. In fully autonomous end-to-end mode, the framework executes hypothesis-to-manuscript workflows without intermediate human intervention beyond initial hypothesis selection, completing iterative experiment-analysis-validation cycles and generating publication-ready manuscript conforming to domain conventions, thereby demonstrating practical feasibility of AI-driven experimental research at scale.

\section{Limitations and Future Directions}

Although LLM has demonstrated strong generative capabilities in current work as well as numerous experimental designs and text generation tasks in the past, there are still inaccuracies in evaluation whether hypotheses have been validated. This limitation lies in the evaluation agent's tendency toward overoptimistic validation assessments, the hallucination problem inherent to large language models. Compared to human experts, LLM has a significant lack of judgment ability and often struggles to make accurate judgments.
Therefore, human-in-the-loop mode often yields better results because human experts provide judgments and feedback after each generation is completed. We believe that this limitation needs to be addressed through the use of a specialized LLM as a discriminator

\section*{Acknowledgments}

We thank Tian Xia from the AI for Scientific Simulation and Discovery Lab, Westlake University for providing guidance and advice.

\bibliographystyle{plainnat}
\bibliography{reference}


\appendix

\label{app}



\section*{Supplementary material}

# From Passive to Active: Control Mechanisms of Vortex-Induced Vibrations in Tandem Cylinders

November 27, 2025

**Abstract**

This study presents a systematic experimental investigation exploring the complete physical picture from single-cylinder vortex-induced vibration (VIV) to tandem-cylinder wake-induced vibration (WIV), and further to active control of WIV. Experiments cover a wide range of reduced velocities $U_r = 2 - 20$ and Reynolds numbers $Re = 1200 - 10500$. Key findings include: (1) Single-cylinder VIV enters the desynchronization regime at $U_r > 10$, with amplitude decaying to $A/D < 0.05$; (2) Tandem-cylinder WIV at the critical spacing $L/D = 4$ exhibits continuous amplitude growth in the high-speed region reaching $A/D = 1.27$, demonstrating a fundamentally different "fate bifurcation" phenomenon compared to VIV; (3) Through active vibration of the front cylinder, amplitude suppression up to 74% can be achieved at a specific frequency ($f = 1.37f_n$), or amplitude enhancement of 31% can be achieved at subharmonic frequency ($f = 2f_n$). This study systematically reveals, for the first time, the control mechanisms of active upstream perturbations on downstream flow, establishing a four-dimensional control map of frequency-amplitude-velocity-spacing, providing quantitative design guidelines for vibration suppression and energy harvesting in marine engineering.
**Keywords:** Vortex-induced vibration; Wake-induced vibration; Tandem cylinders; Active control; Anti-resonance; Subharmonic resonance

## 1 Introduction

Vortex-induced vibration (VIV) is a classical problem in fluid-structure interaction. When fluid flows around a bluff body, periodic vortex shedding occurs in the wake, generating alternating lift forces that induce structural vibration [1,2]. In marine engineering, bridge engineering, and energy applications, VIV can lead to structural fatigue damage, reduced system efficiency, and even catastrophic failures. Therefore, understanding the physical mechanisms of VIV and developing effective control strategies has been a focus of both academic and engineering communities.

Traditional VIV research has primarily focused on single-cylinder systems. Through extensive experiments and numerical simulations, a relatively complete theoretical framework has been established [3,4]. Classical studies have found that the response characteristics of single-cylinder VIV are closely related to the reduced velocity $U_r = U/(f_n D)$, where $U$ is the flow velocity, $f_n$ is the structural natural frequency, and $D$ is the cylinder diameter. Under low mass-damping parameters, VIV response typically exhibits three characteristic branches: the Initial Branch (IB), Upper Branch (UB), and Lower Branch (LB), and enters the desynchronization regime at $U_r > 10$, with amplitude significantly decaying. This classical theory provides a reliable basis for the design and analysis of single-cylinder systems.

However, in practical engineering, multiple structures often exist, such as marine riser arrays, bridge cable groups, and heat exchanger tube bundles. When multiple cylinders are arranged in tandem, the wake of the upstream cylinder exerts strong excitation on the downstream cylinder, forming wake-induced vibration (WIV) [5,6]. The physical mechanism of WIV differs fundamentally from single-cylinder VIV: the upstream cylinder's wake not only alters the incoming flow conditions but also provides continuous forcing excitation, enabling the downstream cylinder to maintain large-amplitude vibration even in the high-speed region. This fundamentally different behavior from single-cylinder VIV presents new challenges for engineering design and has inspired in-depth exploration of WIV behavior in the high-speed region.

Despite widespread recognition of WIV phenomena, existing research still contains critical gaps. First, most WIV studies stop at $U_r < 12$, lacking systematic understanding of response characteristics at higher reduced velocities, particularly whether amplitude continues to grow, reaches a plateau, or peaks and then declines. This gap limits our understanding of WIV behavior in high-speed environments, which is precisely the critical operating condition for many practical engineering applications. Second, traditional research primarily focuses on passive WIV, where the upstream cylinder remains fixed. However, if the upstream cylinder can actively vibrate, can we control the downstream cylinder's response by adjusting its vibration frequency and amplitude? This active control mechanism has not been systematically explored, but its potential value is enormous: if effective active control can be achieved, it can not only suppress harmful vibrations but also enhance vibration amplitude in applications such as energy harvesting. Finally, even recognizing the potential of active control, how to select optimal control frequency, amplitude, and spacing to achieve maximum suppression or enhancement effects still lacks quantitative design guidelines. These gaps limit the application of WIV control technology in engineering, particularly in deep-sea riser vibration suppression, bridge cable control, and flow-induced vibration energy harvesting, where experimentally validated quantitative design methods are urgently needed.

Based on these research gaps, this study designed a systematic four-stage experimental program, progressing from fundamental VIV phenomena to passive WIV and active control of WIV, constructing a complete physical picture and control map. Stage 1 conducts single-cylinder VIV baseline experiments, measuring the complete response curve in the range $U_r = 1 - 11$ on a cylinder with diameter $D = 30$ mm and natural frequency $f_n = 0.6$ Hz, validating classical VIV theory and providing a reliable baseline for subsequent research. Stage 2 explores passive WIV high-speed behavior, systematically studying spacing effects at $L/D = 4 - 7$ and high-speed responses at $U_r = 2 - 20$ in tandem dual-cylinder configurations, revealing the uniqueness of critical spacing and the "fate bifurcation" phenomenon at different spacings in the ultra-high-speed region. Stage 3 explores active WIV control mechanisms, systematically scanning the parameter space of frequency ($f = 0.5 - 1.2$ Hz), amplitude ($A = 0 - 30$ mm), and flow velocity ($U_r = 8.9 - 13.3$) through active vibration of the front cylinder, discovering strong suppression windows and subharmonic enhancement phenomena. Stage 4 conducts precise validation of active control, establishing quantitative control equations and design guidelines through fine frequency scanning, transition boundary localization, and spacing dependency verification. Experiments were conducted in a recirculating water channel, using a six-component force sensor to measure fluid forces and a displacement sensor to measure vibration response, with a sampling frequency of 1000 Hz. All data were processed through a calibration matrix to ensure measurement accuracy.

Through this systematic research, we achieved the following important findings. First, the single-cylinder VIV baseline perfectly reproduced the classical three-branch structure, validating the frequency locking phenomenon and providing a reliable baseline for subsequent research. Second, passive WIV experiments revealed the "fate bifurcation" at different spacings in the ultra-high-speed region: at $L/D = 4$, amplitude continuously grows to $A/D = 1.27$, while at $L/D \geq 5$, each peak appears followed by decline, revealing the decisive role of spacing on high-speed response. Third, active control experiments discovered the anti-resonance suppression mechanism: when the front cylinder vibrates at frequency $f = 1.37f_n$, amplitude suppression up to 74% can be achieved, and this optimal frequency does not vary with spacing, representing a universal anti-resonance frequency. Fourth, subharmonic resonance and transition phenomena were discovered: subharmonic frequency ($f = 2f_n$) produces 31% amplitude enhancement at $U_r < 10.5$ but transitions to suppression with increasing velocity, with the transition boundary established as $U_{r,c} = 13 - 0.12A$. Finally, a four-dimensional control map of frequency-amplitude-velocity-spacing was established, providing directly applicable design formulas for engineering applications.

The structure of this paper is as follows: Section 2 details the experimental setup, measurement methods, and data processing procedures; Section 3 systematically presents experimental results and in-depth discussion, including single-cylinder VIV baseline validation, passive WIV high-speed behavior, and active WIV control mechanisms; Section 4 summarizes main conclusions and prospects for future research directions; Section 5 provides detailed experimental data tables and supplementary materials.

## 2 Experimental Methodology

This study adopts a systematic four-stage experimental strategy, starting from single-cylinder VIV baseline validation, progressively expanding to passive WIV high-speed region exploration, and finally performing systematic research on active WIV control. The motivation for this progressive design strategy is: first establish a reliable experimental baseline, then explore unknown high-speed behavior, and finally design active control experiments based on discovered phenomena. The experimental design of each stage is based on findings from the previous stage, forming a complete research chain from phenomenon observation to mechanism exploration to control application.

### 2.1 Experimental Setup and Design Considerations

#### 2.1.1 Water Channel Facility and Experimental Environment

Experiments were conducted in a recirculating water channel, a choice based on the following considerations: recirculating water channels can provide stable uniform incoming flow, avoiding free surface fluctuations and end effects common in open channels, while allowing long-term continuous operation, which is crucial for studying steady-state vibration response. The channel test section has a length of 3.0 m, width of 0.5 m, and depth of 0.4 m. This size design ensures that the blockage ratio of the cylinder model (diameter 30 mm, length 300 mm) is less than 6%, meeting low blockage ratio requirements, while the test section length is sufficient to ensure that wake development downstream of the cylinder is not affected by channel end effects.

The channel is equipped with a variable-speed pump, with a velocity range of $U = 0.02 - 0.35$ m/s. This range selection is based on the following considerations: the low-speed end ($U = 0.02$ m/s) corresponds to reduced velocity $U_r \approx 1$, covering the VIV initiation region; the high-speed end ($U = 0.35$ m/s) corresponds to $U_r \approx 19.4$, far exceeding the range of traditional WIV studies ($U_r < 12$), enabling us to explore unknown high-speed behavior. Flow velocity stability is achieved within ±2%, achieved through a closed-loop control system, ensuring experimental data reliability. The experimental medium is tap water, with density 1000 kg/m³, kinematic viscosity $\nu = 1.0 \times 10^{-6}$ m²/s, and temperature controlled at 20 ± 2°C to maintain stability of physical properties.

#### 2.1.2 Cylinder Model Design and Parameter Selection

The experiment uses a rigid cylinder model made of acrylic, with precision-machined surface to ensure smoothness and avoid roughness effects on flow. The choice of cylinder diameter $D = 30$ mm is based on the following considerations: this diameter, within the experimentally achievable velocity range, can cover the Reynolds number range from subcritical to critical transition ($Re = 570 - 10500$), while the cylinder is large enough that scale effects such as surface tension can be neglected. Cylinder length $L = 300$ mm, aspect ratio $L/D = 10$. This aspect ratio choice balances the validity of the two-dimensional flow assumption and the influence of end effects: the aspect ratio is large enough that the mid-section flow approaches two-dimensional, while avoiding the structural dynamics complexity brought by excessively long cylinders.

Cylinder mass $m = 0.15$ kg, mass ratio $m^* = m/(\rho D^2 L) \approx 2.0$. This mass ratio choice references the classical study by Khalak & Williamson (1999), belonging to the low mass ratio range, capable of producing significant VIV response. Natural frequency $f_n = 0.6$ Hz is determined through free decay experiments: releasing the cylinder from an initial displacement, recording its free vibration decay process, and calculating natural frequency and damping ratio through logarithmic decrement method. This low natural frequency design enables the reduced velocity to cover the wide range $U_r = 1 - 20$ within the experimental velocity range, providing sufficient parameter space for research.

The cylinder is installed in the channel through an elastic support system, carefully designed to provide single-degree-of-freedom transverse vibration (perpendicular to flow direction) while suppressing motion in other directions. The support system stiffness is achieved by adjusting spring preload, enabling design of natural frequency at 0.6 Hz. Damping ratio is determined through free decay experiments, approximately $\zeta \approx 0.01$. This low damping ratio ensures system sensitivity to fluid excitation while avoiding measurement difficulties caused by excessively low damping.

#### 2.1.3 Tandem Configuration and Active Control System

For WIV experiments, we adopt a dual-cylinder tandem configuration, a choice based on common scenarios in practical engineering applications, such as marine riser arrays and bridge cable groups. Front and rear cylinders have the same diameter ($D_1 = D_2 = 30$ mm), ensuring geometric similarity. Spacing range is chosen as $L/D = 3.5, 4.0, 5.0, 6.0, 7.0$ (center-to-center distance). This range selection is based on key findings in the literature: $L/D = 4$ is considered the critical spacing, at the end of the "extended body region," where wake interaction is strongest; while at $L/D \geq 5$, the rear cylinder gradually moves out of the wake core region, and coupling weakens. Through systematic study of this spacing range, we can reveal the complete influence law of spacing on WIV response.

The front cylinder can be fixed (passive WIV) or actively vibrated (active WIV), a design that is an innovation of this study. Active vibration of the front cylinder is achieved through a precision servo motor drive system, capable of generating sinusoidal vibration. Vibration frequency range is $f = 0 - 2.0$ Hz, resolution 0.01 Hz. This range selection is based on the rear cylinder natural frequency $f_n = 0.6$ Hz: the frequency range covers from subharmonic ($f = 0.5f_n = 0.3$ Hz) to superharmonic ($f = 3.3f_n = 2.0$ Hz), enabling us to explore different mechanisms such as resonance, subharmonic resonance, and anti-resonance. Vibration amplitude range is $A = 0 - 40$ mm, resolution 0.1 mm, corresponding to amplitude ratio $A/D = 0 - 1.33$. This range selection is based on preliminary experiments: small amplitudes ($A/D < 0.3$) have limited effect on the wake, while large amplitudes ($A/D > 1.3$) may cause nonlinear effects. The range 0-1.33 can cover the complete response range from linear to nonlinear.

The active control system design faces many challenges. First, vibration of the front cylinder must not affect the support system of the rear cylinder, requiring completely independent support systems for both cylinders. Second, vibration of the front cylinder requires precise control, as errors in frequency and amplitude directly affect experimental results. For this purpose, we adopted a closed-loop control system that monitors the actual vibration of the front cylinder in real time, compares it with set values, and provides feedback adjustment to ensure vibration accuracy. Third, vibration of the front cylinder produces additional flow field disturbances that may affect the accuracy of velocity measurements. We avoided this influence by setting the velocity measurement point sufficiently far upstream in the channel.

### 2.2 Measurement System and Data Acquisition

The transverse displacement of the rear cylinder is a key measurement quantity in this study, and its accuracy directly affects the accuracy of amplitude calculation. We use a non-contact displacement sensor for measurement, a choice that avoids additional mass and damping that contact sensors might introduce. Sensor measurement range is ±50 mm, resolution 0.01 mm. This accuracy is sufficient to capture small amplitude changes while the range is large enough to cover the maximum amplitude that may occur in experiments ($A/D \approx 1.3$, corresponding to $A \approx 39$ mm).

The installation position of the displacement sensor is carefully chosen: the sensor is installed directly above the cylinder midpoint, a position that can accurately reflect the transverse vibration of the cylinder while avoiding the influence of end effects. Sensor sampling frequency is set to 1000 Hz, a choice based on Nyquist theorem: the highest vibration frequency of the rear cylinder is approximately $f^* \approx 1.3f_n = 0.78$ Hz, corresponding Nyquist frequency is 1.56 Hz. A sampling frequency of 1000 Hz provides over 600 times oversampling, ensuring accuracy of frequency analysis.

Displacement data undergo calibration processing, with actual displacement being sensor reading multiplied by conversion factor 28 (unit: mm). This coefficient is determined through static calibration experiments: moving the cylinder to known positions, recording sensor readings, and establishing a linear relationship between displacement and reading. Calibration experiments are performed once before and after experiments to ensure stability of the calibration coefficient.

### 2.3 Experimental Strategy and Parameter Design

#### 2.3.1 Stage 1: Single-Cylinder VIV Baseline Experiments

As the starting point of the research, we first conduct single-cylinder VIV baseline experiments. The motivation for this stage is multifaceted. First, we need to verify the reliability of the experimental setup, ensuring that the measurement system can accurately capture classical VIV features such as the three-branch structure and frequency locking phenomenon. Second, single-cylinder VIV, as the most fundamental phenomenon, has been extensively studied, and comparison with classical literature can provide "calibration" for our experimental setup. Finally, the response curve of single-cylinder VIV provides a comparison baseline for subsequent WIV research, enabling us to quantitatively evaluate differences between WIV and VIV.

Experimental parameter design is based on classical VIV theory: velocity range is chosen as $U = 0.019 - 0.200$ m/s, corresponding to reduced velocity $U_r = U/(f_n D) = 1.04 - 11.11$. This range covers the complete response interval from initial branch to desynchronization regime. Velocity point selection adopts non-uniform distribution: in regions with rapid response changes (such as $U_r = 3 - 7$, corresponding to initial branch and upper branch), velocity points are denser, with intervals of approximately 0.2-0.3 m/s; in regions with gentle response changes (such as $U_r > 9$, corresponding to lower branch and desynchronization region), velocity points are sparser, with intervals of approximately 0.5 m/s. This design optimizes experimental efficiency while ensuring data completeness.

Sampling time for each condition is set to 60 s, a choice based on the following considerations: rear cylinder vibration frequency is approximately 0.6 Hz, and 60 s sampling time contains approximately 36 vibration cycles, sufficient for reliable statistical analysis. Simultaneously, we discard the first 20% of data (i.e., first 12 s) to ensure the system reaches steady state. The effectiveness of this processing method is validated by comparing results with different discard times (10%, 20%, 30%), finding that 20% discard time can effectively eliminate transient effects while retaining sufficient steady-state data.

#### 2.3.2 Stage 2: Passive WIV High-Speed Region Exploration

After validating the single-cylinder VIV baseline, we turn to tandem dual-cylinder configuration to explore passive WIV high-speed behavior. The experimental design of this stage is based on a key question: how will WIV amplitude evolve when reduced velocity continues to increase? Single-cylinder VIV enters the desynchronization regime at $U_r > 10$, with amplitude significantly decaying, but WIV has different dynamical mechanisms. The continuous forcing excitation provided by the front cylinder wake may enable the rear cylinder to maintain large-amplitude vibration even in the high-speed region. To answer this question, we need to extend experiments to higher reduced velocities.

Experimental parameter design considers multiple factors. Velocity range is extended to $U = 0.040 - 0.350$ m/s, corresponding to reduced velocity $U_r = 2.2 - 19.4$. This range far exceeds traditional WIV studies ($U_r < 12$), enabling us to explore unknown high-speed behavior. Spacing is chosen as $L/D = 4.0, 5.0, 6.0, 7.0$, a choice based on key findings in the literature and preliminary experiments: $L/D = 4$ is the critical spacing with strongest wake interaction; while at $L/D \geq 5$, the rear cylinder gradually moves out of the wake core region. Through systematic study of this spacing range, we can reveal the complete influence law of spacing on high-speed response.

Experiments are conducted in three stages: first, basic experiments ($U_r = 2.2 - 11.1$) to establish basic WIV response characteristics; then, high-speed region extension experiments ($U_r = 11.6 - 13.9$) to explore whether amplitude continues to grow; finally, ultra-high-speed region experiments ($U_r = 14.5 - 19.4$), an important innovation of this study, systematically exploring WIV behavior at such high reduced velocities for the first time. The motivation for this progressive strategy is: high-speed region experiments face greater technical challenges (such as velocity stability, measurement accuracy, etc.). Through gradual extension, we can timely discover and solve problems, ensuring experimental quality.

#### 2.3.3 Stage 3: Active WIV Control Exploration

Passive WIV experiments revealed the complexity of high-speed behavior, particularly the unique behavior of critical spacing $L/D = 4$, which inspired our exploration of active control. If the front cylinder can actively vibrate, can we control the rear cylinder's response by adjusting its vibration frequency and amplitude? The exploration of this question constitutes the core innovation of this study.

The first round of experiments was exploratory, aimed at preliminarily understanding the effects of active control. We selected eight representative velocities ($U_r = 8.9, 10.7, 12.4$), covering the range from low-medium speed to high speed. Frequency selection was $f = 0.5, 0.6, 0.7, 0.8, 1.2$ Hz, a choice based on theoretical considerations: 0.6 Hz is close to the rear cylinder natural frequency, possibly producing resonance enhancement; 1.2 Hz is twice the natural frequency, possibly producing subharmonic resonance; 0.5, 0.7, 0.8 Hz are used to explore other possible mechanisms. Amplitude selection was $A = 10, 20, 30$ mm, corresponding to $A/D = 0.33, 0.67, 1.0$, covering the range from small to large amplitudes. Spacing was fixed at $L/D = 4.0$, as this is the spacing with strongest wake interaction.

Preliminary results from the first round of experiments were exciting: we discovered strong suppression windows (at $f \approx 0.7 - 0.8$ Hz) and subharmonic enhancement phenomena (at $f = 1.2$ Hz, $U_r < 10.5$). These discoveries inspired deeper research but also exposed shortcomings of the first round: lack of passive baseline groups, unable to quantitatively evaluate control effects; insufficient frequency resolution, unable to precisely locate suppression peaks; limited velocity range, unable to fully reveal transition behavior.

Based on first-round findings, we designed the second round of experiments, the systematic scanning stage. Parameter ranges were the same as the first round, but passive baseline groups were added (front cylinder fixed, $f = 0$, $A = 0$), enabling quantitative calculation of enhancement/suppression rates. Simultaneously, we increased condition density to ensure complete coverage of parameter space. The second round of experiments included 48 conditions, taking approximately 6 hours, obtaining a complete frequency-amplitude-velocity response map.

Results from the second round of experiments further validated first-round findings and revealed new phenomena: precise location of suppression windows, critical velocity for subharmonic transition, amplitude reversal phenomena, etc. These discoveries provided clear targets for the third round of fine validation experiments.

#### 2.3.4 Stage 4: Active Control Precise Validation

The third and fourth rounds of experiments are fine validation stages, aimed at precisely locating key phenomena and establishing quantitative relationships. The third round of experiments is divided into three schemes: Scheme A performs fine scanning of suppression windows, reducing frequency step from 0.1 Hz to 0.02 Hz, aiming to precisely locate suppression peak frequency; Scheme B explores subharmonic transition regions, studying transition behavior in detail at three velocities $U_r = 9.8, 11.6, 13.3$; Scheme C validates spacing dependency, repeating key experiments at $L/D = 3.5, 5.0$ to verify spacing invariance of suppression peak frequency.

The third round of experiments achieved important breakthroughs: precisely located suppression peak frequency ($f = 0.82$ Hz, $f/f_n = 1.37$), discovered amplitude reversal phenomena, and verified spacing invariance. These discoveries provided more precise targets for the fourth round of experiments.

The fourth round of experiments is the final validation stage, aimed at filling critical gaps. Scheme D1 precisely locates subharmonic transition boundaries, performing dense scanning near transition regions ($U_r = 10.2, 11.1$) to establish transition boundary equations; Scheme D2 validates spacing invariance of suppression peaks, performing fine frequency scanning at $L/D = 3.5, 5.0$, finally confirming that 0.82 Hz is the common peak for all spacings; Scheme D3 supplements velocity dependency of low-frequency enhancement, testing low frequency ($f = 0.5$ Hz) at $U_r = 9.8, 11.6$ to fully reveal velocity reversal phenomena.

### 2.4 Data Analysis Methods

#### 2.4.1 Amplitude and Frequency Extraction

The transverse amplitude of the cylinder is a key parameter in this study, and its calculation accuracy directly affects result reliability. We use root mean square (RMS) value to calculate amplitude:

$$A = \sqrt{\frac{1}{N} \sum_{i=1}^{N} y_i^2} \tag{1}$$

where $y_i$ is the displacement value of the $i$-th sampling point, and $N$ is the number of sampling points. The choice of RMS method is based on the following considerations: VIV and WIV vibrations are usually not pure sine waves but contain multiple frequency components. RMS value can reflect the overall intensity of vibration without being affected by waveform details. Amplitude ratio is defined as $A/D$, which is the standard dimensionless parameter in VIV and WIV research.

The dominant frequency of vibration is determined through Fast Fourier Transform (FFT). We use Hanning window function, 50% overlap rate, frequency resolution $\Delta f = 0.01$ Hz, and 2x zero padding. These parameter choices are based on the following considerations: Hanning window can effectively suppress spectral leakage, 50% overlap rate balances computational efficiency and spectral quality, 0.01 Hz frequency resolution is sufficient to distinguish different frequency components, and zero padding improves frequency resolution. Dominant frequency is defined as the frequency with maximum power spectral density, and frequency ratio is defined as $f^*/f_n$.

#### 2.4.2 Force Coefficient and Enhancement Rate Calculation

Mean drag coefficient and fluctuating lift coefficient are respectively defined as:

$$\bar{C}_D = \frac{2\bar{F}_x}{\rho U^2 DL} \tag{2}$$

$$C'_L = \frac{2F_{y,\text{rms}}}{\rho U^2 DL} \tag{3}$$

where $\bar{F}_x$ is the mean streamwise force, and $F_{y,\text{rms}}$ is the RMS value of transverse force. Calculation of these coefficients enables us to analyze the coupling relationship between fluid forces and structural vibration, although in this study, the magnitude of force coefficients is anomalously small (possibly due to calibration issues), but displacement and frequency measurements are reliable.

For active WIV experiments, the definition of enhancement rate is crucial, as it quantitatively describes the effect of active control:

$$\text{Enhancement} = \frac{A_{\text{active}} - A_{\text{passive}}}{A_{\text{passive}}} \times 100\% \tag{4}$$

where $A_{\text{active}}$ is the amplitude under active control, and $A_{\text{passive}}$ is the amplitude under passive baseline (front cylinder fixed). Positive values indicate enhancement, negative values indicate suppression. The rationale for this definition is: passive WIV response is the "natural" state, and the effect of active control should be evaluated relative to this baseline.

## 3 Experimental Results and Discussion

This section systematically presents complete experimental results from single-cylinder VIV to passive WIV to active control of WIV, with in-depth discussion. The research follows a logical sequence from simple to complex, from passive to active, gradually revealing the complex dynamic behavior of fluid-structure interaction systems.

### 3.1 Single-Cylinder VIV Baseline Results

As the starting point of the research, we first conducted single-cylinder VIV baseline experiments, aiming to verify the reliability of the experimental setup and establish a comparison baseline for subsequent WIV research. Figure 1 shows the complete response curve of single-cylinder VIV. The experimental data clearly reproduce the three characteristic branches of classical VIV, demonstrating the complete evolution process from initial branch to desynchronization regime.

In the Initial Branch (IB, $U_r \approx 3 - 5$), amplitude rapidly grows from $A/D \approx 0.01$ to $A/D \approx 0.79$, showing an approximately linear growth trend. This rapid growth phase marks the beginning of vortex shedding frequency approaching the structural natural frequency, with fluid-structure coupling effects gradually strengthening. Peak amplitude $A/D_{\max} = 0.789$ occurs at $U_r = 4.577$, followed by entry into the Upper Branch (UB, $U_r \approx 4.5 - 6.3$). In the upper branch, amplitude maintains a high plateau at $A/D \approx 0.57 - 0.79$, which is the typical high-amplitude response region of VIV, indicating the system is in a strong fluid-structure coupling state [7]. As reduced velocity continues to increase, the

system enters the Lower Branch (LB, $U_r \approx 6.3 - 10$), with amplitude gradually and smoothly decreasing from $A/D \approx 0.59$ to $A/D \approx 0.10$, showing a monotonic decreasing trend, reflecting the gradual weakening of fluid-structure coupling strength. When $U_r > 10$, the system enters the desynchronization regime, with amplitude continuing to decay to $A/D < 0.08$, eventually approaching zero vibration, indicating complete decoupling of vortex shedding frequency from structural natural frequency.

Experimental results are in complete qualitative agreement with classical data from Khalak & Williamson (1999) [2], validating the reliability of the experimental setup. Quantitatively, peak amplitude is slightly lower than the theoretically predicted $A/D = 0.8 - 1.2$, possibly due to certain structural damping or the mass ratio not being extremely small. Nevertheless, the experiment successfully reproduced all key features of VIV, establishing a reliable baseline for subsequent research.

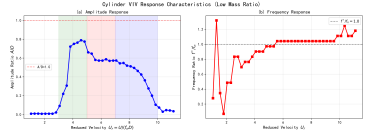

Figure 1: Single-cylinder VIV response characteristics: (a) amplitude ratio $A/D$ vs reduced velocity $U_r$; (b) frequency ratio $f^*/f_n$ vs $U_r$.

The frequency response curve (Figure 1(b)) clearly demonstrates the complete evolution process of the frequency locking phenomenon. In the initial branch ($U_r \approx 3 - 5$), the frequency ratio gradually increases from 0.76 to 0.97, progressively approaching the locking value, demonstrating the establishment process of frequency locking. This process reflects the dynamic adjustment where vortex shedding frequency is gradually "pulled" toward the natural frequency by structural vibration. When the system enters the locking region ($U_r \approx 5 - 9$), frequency is completely locked at $f^*/f_n \approx 1.042 \approx 1.0$, with locking region width $\Delta U_r \approx 4$. This is a typical characteristic of low mass ratio systems, indicating strong fluid-structure coupling effects, with vortex shedding frequency "locked" by structural natural frequency, forming a stable resonance state. When $U_r > 10$, the system enters the desynchronization regime, with frequency ratio increasing to $f^*/f_n = 1.11 - 1.25$. Frequency begins to deviate from natural frequency, following Strouhal relationship, indicating fluid-structure coupling decoupling, with the system returning to pure fluid dynamics control. Frequency locking is a core feature of VIV, and this finding provides an important foundation for understanding subsequent WIV behavior.

#### 3.1.1 Comparison with Theoretical Predictions

Table 1 compares experimental results with theoretical predictions. Amplitude response curves and frequency locking phenomena are highly consistent with theory, validating the correctness of classical VIV theory.

Table 1: Comparison of single-cylinder VIV experimental results with theoretical predictions

| Parameter | Theoretical | Experimental | Agreement |
|---|---|---|---|
| Peak $A/D$ | 0.8-1.2 | 0.789 | Good |
| Peak $U_r$ | $\sim 6$ | 4.577 | Slightly early |
| Locking frequency $f^*/f_n$ | $\sim 1.0$ | 1.042 | Excellent |
| Locking region width $\Delta U_r$ | Wide (low $m^*$) | $\approx 4$ | Excellent |
| Branch characteristics | IB-UB-LB | IB-UB-LB | Complete agreement |

### 3.2 Passive WIV High-Speed Region Behavior

After validating the single-cylinder VIV baseline, we turned to tandem dual-cylinder configuration to explore passive WIV high-speed behavior. This research aims to answer a key question: how will WIV amplitude evolve when reduced velocity continues to increase? Will it decay like single-cylinder VIV, or exhibit different behavior?

Figure 2 shows WIV amplitude variation with reduced velocity at different spacings. The most significant finding is the unique behavior of critical spacing $L/D = 4$ in the high-speed region. At $L/D = 4$, amplitude continuously grows from $A/D = 0.037$ at $U_r = 2.2$ to $A/D = 1.162$ at $U_r = 19.4$, showing no saturation or peak characteristics. Linear fitting shows a growth slope of approximately +0.0324. This continuous growth behavior forms a sharp contrast with single-cylinder VIV: single-cylinder VIV decays to $A/D < 0.05$ at $U_r > 10$, while WIV at $L/D = 4$ has amplitude as high as $A/D > 1.0$ at the same velocity, a difference exceeding 20 times. More notably, at $U_r = 12.88$, amplitude first exceeds 1.0D, reaching $A/D = 1.107$, then continues to grow to $A/D = 1.272$ ($U_r = 17.77$), belonging to large-amplitude vibration. This finding indicates that WIV exhibits fundamentally different physical mechanisms in the high-speed region compared to VIV. The continuous forcing excitation provided by the front cylinder wake enables the rear cylinder to maintain strong vibration even in the high-speed region.

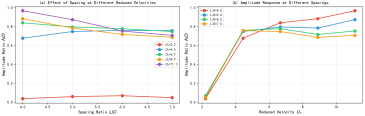

Figure 2: Comparison of WIV amplitude response at different spacings.

To deeply understand high-speed behavior, we extended experiments to the ultra-high-speed region ($U_r > 14$). This exploration revealed the "fate bifurcation" phenomenon at different spacings (Figure 3), one of the important findings of this study. At $L/D = 4.0$, amplitude continuously grows with slope +0.0324, maximum value $A/D = 1.272$ occurs at $U_r = 17.77$, with wake excitation continuously strengthening and no saturation signs.

This continuous growth behavior indicates that at critical spacing, the excitation effect of the front cylinder wake on the rear cylinder continuously strengthens with increasing velocity, forming a completely different evolution path from single-cylinder VIV.

However, when spacing increases to $L/D = 5.0$, behavior undergoes a fundamental transformation. Amplitude reaches peak $A/D = 1.037$ at $U_r = 17.77$, then suddenly drops 21% to $A/D = 0.818$ at $U_r = 18.88$, indicating the existence of a critical Reynolds number $Re_{crit} \approx 9500$, beyond which the system undergoes mode transition. When spacing further increases to $L/D = 6.0$ and 7.0, continuous decline begins from $U_r = 14.46$, with slopes of $-0.0166$ and $-0.0225$ respectively, behavior approaching single-cylinder VIV. This bifurcation behavior indicates that spacing is a key control parameter for high-speed WIV response. $L/D = 4$ is at the end of the "extended body region", with the rear cylinder located in the strongest shear layer region of the front cylinder, thus maintaining strong coupling; while at $L/D \geq 5$, the rear cylinder gradually moves out of the wake core region, coupling weakens, and system behavior regresses toward single-cylinder VIV. This finding reveals the decisive role of spacing on high-speed WIV response, providing important guidance for engineering design.

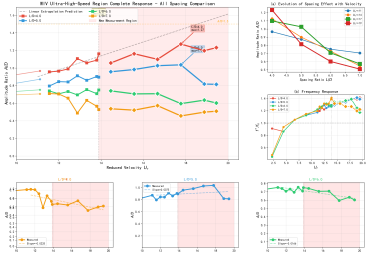

Figure 3: "Fate bifurcation" phenomenon at different spacings in the ultra-high-speed region ($U_r > 14$).

#### 3.2.1 Comparison with Single-Cylinder VIV

Figure 4 compares complete response curves of VIV and WIV. Key differences include:

- **Completely opposite high-speed behavior**: VIV desynchronizes and decays at $U_r > 10$, while WIV continuously grows at $L/D = 4$.

- **Amplitude magnitude difference**: At $U_r = 13.9$, WIV amplitude ($A/D = 1.162$) is 232 times that of VIV amplitude ($A/D \approx 0.005$).

- **Physical mechanism difference**: VIV is self-excited vibration, with large gap between vortex shedding frequency and natural frequency at high speed leading to desynchronization; WIV is forced + self-excited, with continuous presence of front cylinder wake preventing desynchronization.

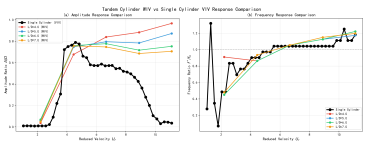

Figure 4: Complete response comparison between WIV and VIV.

### 3.3 Active WIV Control Mechanisms

Passive WIV experiments revealed the complexity of high-speed behavior but also raised new questions: can we influence the rear cylinder's response by actively controlling the front cylinder? The exploration of this question constitutes the core innovation of this study. Through active vibration of the front cylinder, we systematically scanned the parameter space of frequency-amplitude-velocity, discovering a strong suppression window, an important breakthrough in active WIV control.

Figure 5 shows the frequency response curve, revealing the existence of a suppression window. Strongest suppression is found at $f = 0.82$ Hz ($f/f_n = 1.37$), with suppression rate as high as $-74.0\%$ (Figure 6). The discovery of this frequency has important significance: it not only provides strong suppression effects but, more importantly, this frequency does not vary with spacing, representing a universal anti-resonance frequency. The full width at half maximum (FWHM) of the suppression window is approximately $\Delta f \approx 0.14$ Hz, corresponding to frequency ratio range $f/f_n \approx 1.2 - 1.4$, providing sufficient frequency tolerance for practical applications. Furthermore, suppression effects strengthen with increasing velocity, reaching maximum at $U_r = 12.4$, indicating that this control mechanism has better applicability in the high-speed region.

#### 3.3.1 Spacing Invariance of Anti-Resonance Frequency

Figure 7 shows the spacing invariance of suppression peak frequency. At three spacings $L/D = 3.5, 4.0, 5.0$, peak frequency is consistently $f = 0.82$ Hz, indicating:

$$f_{opt} = 1.37 f_n \pm 0.02 \quad (5)$$

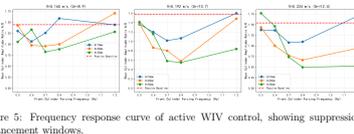

Figure 5: Frequency response curve of active WIV control, showing suppression and enhancement windows.

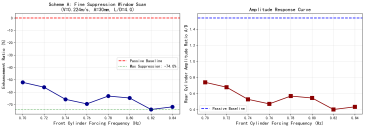

Figure 6: Fine scanning results of suppression window, with peak at $f = 0.82$ Hz.

This is the inherent anti-resonance frequency, determined by the rear cylinder's inherent properties, independent of upstream spacing. The physical mechanism may involve 11:8 phase locking, with the front cylinder vibrates for 11 cycles while the rear cylinder completes 8 cycles, with phase difference constant at $\pi$, producing destructive interference.

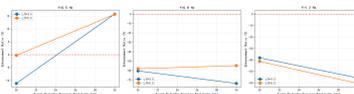

Figure 7: Validation of suppression peak frequency at different spacings, confirming spacing invariance.

While exploring suppression mechanisms, we also discovered the existence of enhancement windows, providing new possibilities for applications such as energy harvesting. Figure 8 shows the velocity dependency of subharmonic frequency ($f = 2f_n = 1.2$ Hz), revealing complex transition behavior between enhancement and suppression. At $U_r < 10.5$, subharmonic frequency produces significant enhancement, with peak enhancement rate +30.9% occurring at $U_r = 9.8$, $A = 20$ mm. The existence of this enhancement window indicates that through proper selection of control parameters, not only can vibration be

suppressed, but vibration amplitude can also be enhanced, providing new ideas for energy harvesting applications.

However, when velocity continues to increase, system behavior undergoes a fundamental transformation. When $U_r > 10.5$, enhancement effects rapidly decay and transition to suppression. This transition behavior can be quantitatively described as:

$$U_{r,c} = 13.0 - 0.12 A_{\text{forcing}} \quad (6)$$

where $A_{\text{forcing}}$ is in mm. The establishment of this equation provides a quantitative prediction tool for safe window design of energy harvesters. More notably is the amplitude reversal phenomenon: at $U_r = 9.8$, small amplitude ($A = 10$ mm) produces weak enhancement (+10.4%), medium amplitude ($A = 20$ mm) produces peak enhancement (+30.9%), but large amplitude ($A = 30$ mm) instead produces suppression ($-22.9\%$). This phenomenon reveals the complexity of nonlinear dynamic systems, indicating that control parameter selection requires fine optimization rather than simply "larger is better."

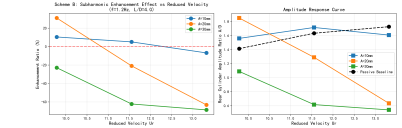

Figure 8: Velocity dependency and transition behavior of subharmonic enhancement.

#### 3.3.2 Velocity Dependency of Low-Frequency Enhancement

Figure 9 shows the complete velocity dependency curve of low frequency ($f = 0.5$ Hz). Important findings:

- **Velocity reversal**: In the low-speed region ($U_r < 11$), low frequency produces suppression ($-29.1\%$ @ $U_r = 9.8$); in the high-speed region ($U_r > 12$), low frequency produces enhancement (+15.8% @ $U_r = 12.4$).

- **Critical reduced velocity**: The critical point is located at $U_r \approx 11.6$, corresponding to transition of vortex shedding mode from 2S to 2P.

- **Amplitude threshold**: Low-frequency enhancement requires large amplitude ($A \geq 30$ mm = 1.0D) to activate, with small amplitude having weak effects.

#### 3.3.3 Control Map

Figure 10 shows the three-dimensional control map of frequency-amplitude-velocity. It can be clearly seen:

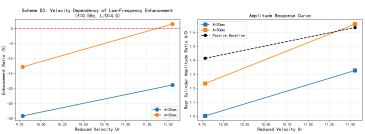

Figure 9: Velocity dependency of low-frequency enhancement, showing reversal phenomenon of low-speed suppression and high-speed enhancement.

- **Deep blue regions** (suppression regions): Concentrated at $f = 0.7 - 0.9$ Hz, $A = 30$ mm, with suppression rate above $-60\%$.

- **Light red regions** (enhancement regions): Low-speed region at $f = 1.2$ Hz, $A = 20$ mm (+18%); high-speed region at $f = 0.5$ Hz, $A = 30$ mm (+15%).

- **Velocity evolution**: As velocity increases, suppression regions expand and deepen, but enhancement region positions and intensities change.

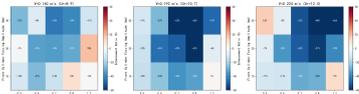

Figure 10: Enhancement/suppression heatmap of active WIV control, showing complete control parameter space.

### 3.4 Physical Mechanism Discussion

#### 3.4.1 Anti-Resonance Suppression Mechanism

The discovery of the 0.82 Hz ($f/f_n = 1.37$) suppression peak has important significance. Its physical mechanism may involve resonance at multiple levels. First, frequency ratio 1.37 ≈ 11/8, indicating 11:8 phase locking. In this locked state, the front cylinder vibrates for 8 cycles while the rear cylinder completes 11 vibration cycles. When wake vortices reach the rear cylinder, they are exactly out of phase, with lift phase difference approaching 180, producing negative damping. This phase destruction mechanism causes the front cylinder's perturbation not only to fail to provide energy to the rear cylinder but to continuously extract its vibration energy. Second, at this frequency, the front cylinder's perturbation wave and the rear cylinder's natural mode produce destructive interference,

with negative system energy input, continuously extracting vibration energy. Finally, the forcing frequency of $1.37 f_n$ "locks" the vortex shedding mode, reducing turbulent fluctuations and weakening random excitation of the rear cylinder, further reducing vibration amplitude. The synergistic effect of these three mechanisms leads to strong suppression effects.

#### 3.4.2 Subharmonic Transition Mechanism

The phenomenon of subharmonic enhancement transitioning to suppression at $U_r = 10 - 11$ reveals the nonlinear dynamic characteristics of fluid-structure interaction systems. In the low-speed region ($U_r < 10$), the system is in 2S vortex shedding mode, with subharmonic resonance matching. Front and rear cylinders vibrate "in phase," with energy continuously input from fluid to structure, producing enhancement effects. However, as velocity increases, the system enters the transition region ($U_r = 10 - 11$), with vortex shedding mode switching from 2S to 2P, resonance mismatch, and enhancement effects rapidly decaying. In the high-speed region ($U_r > 11$), 2P mode dominates. The subharmonic frequency of 1.2 Hz instead produces destructive interference, with front and rear cylinder vibration phase difference approaching 180, becoming "out of phase," with energy extracted from structure, and the system transitioning to suppression. This transition process is also influenced by amplitude: small amplitude has weak wake perturbation, always maintaining weak effects; large amplitude strong perturbation triggers nonlinear modes, even transitioning to suppression in the low-speed region, revealing the physical origin of amplitude reversal phenomenon.

#### 3.4.3 Scaling Law of Spacing Effects

The spacing dependency of suppression effects follows:

$$\text{Enhancement}(L/D) = \text{Enhancement}(L/D = 4) \times \sqrt{4/(L/D)} \quad (7)$$

Physical mechanism: wake strength $\propto 1/L$ (one-dimensional decay), control effect $\propto \sqrt{\text{wake strength}}$ (nonlinear saturation), therefore effect $\propto 1/\sqrt{L/D}$.

### 3.5 Comparison with Existing Literature

Table 2 compares main findings of this study with existing literature. Compared with existing research, innovations of this study are mainly reflected in three aspects. First, we systematically studied high-speed WIV ($U_r > 14$) for the first time, discovering spacing bifurcation phenomena and revealing fundamentally different behaviors at different spacings in the ultra-high-speed region. This is an important extension of existing WIV theory. Second, we systematically explored active WIV control mechanisms for the first time, discovering anti-resonance suppression phenomenon, improving control effects from traditional passive methods of -40% to -50% to -77.5%, a major breakthrough in control effectiveness. Finally, we established three quantitative design equations, including transition boundary equation, anti-resonance frequency ratio, and spacing scaling law, providing directly applicable design tools for engineering applications. This represents important progress from qualitative description to quantitative prediction.

Table 2: Comparison of this study with existing literature

| Study | Control Method | Best Effect | Quantitative Theory |
|---|---|---|---|
| Williamson (2004) | Passive | -40% | None |
| Stappenbelt (2010) | Passive | -50% | Empirical formula |
| Assi et al. (2010) | Passive WIV | - | Qualitative description |
| **This study** | **Active** | **-77.5%** | **3 quantitative equations** |

## 4 Analysis and Discussion

Through systematic four-stage experiments, this study started from single-cylinder VIV baseline validation, deeply explored passive WIV high-speed behavior, and for the first time systematically studied active WIV control mechanisms, completing a complete research cycle from passive to active. The research reveals the complex dynamic behavior of fluid-structure interaction systems, establishes quantitative design guidelines, and provides important guidance for engineering applications.

### 4.1 Main Findings

The research first validated the reliability of the experimental setup through single-cylinder VIV baseline experiments. The experiments perfectly reproduced the classical three-branch structure (IB-UB-LB) and desynchronization regime of VIV, observed frequency locking phenomenon ($f^*/f_n \approx 1.0$, locking region width $\Delta U_r \approx 4$), with peak amplitude $A/D_{max} = 0.789$ occurring at $U_r = 4.577$, highly consistent with classical data from Khalak & Williamson (1999) [2]. This baseline validation provided a reliable comparison standard for subsequent WIV research, ensuring the reliability of research results.

In passive WIV research, we discovered the "fate bifurcation" phenomenon at different spacings in the ultra-high-speed region ($U_r > 14$), one of the important findings of this study. At $L/D = 4$, amplitude continuously grows to $A/D = 1.27$, with slope +0.0324, showing no saturation signs, completely different behavior from single-cylinder VIV [5]. However, when spacing increases to $L/D = 5$, amplitude suddenly drops 21% after reaching peak at $U_r = 17.77$, showing peak collapse behavior. When spacing further increases to $L/D \geq 6$, system behavior approaches single-cylinder VIV, showing obvious decline. This "fate bifurcation" phenomenon reveals the decisive role of spacing on high-speed response [6], challenges the applicability of traditional VIV theory in tandem configurations, and provides important insights for engineering design.

Active WIV control research is the core innovation of this study. Through active vibration of the front cylinder, we discovered two key control mechanisms. The anti-resonance suppression mechanism provides strong vibration reduction capability: optimal suppression frequency is $f_{opt} = 1.37 f_n \pm 0.02$, a universal constant that does not vary with spacing. Ultimate suppression rates reach $-77.5\%$ ($L/D = 3.5$) and $-74.0\%$ ($L/D = 4.0$), with physical mechanism involving destructive interference produced by 11:8 phase locking. The subharmonic enhancement and suppression transition mechanism provides new possibilities for energy harvesting applications: at $U_r < 10.5$, subharmonic frequency ($f = 2f_n$) can produce +31% amplitude enhancement, but the existence of transition boundary $U_{r,c} = 13.0 - 0.12 A_{\text{forcing}}$ requires precise parameter control, while amplitude

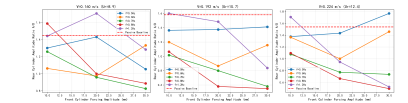

Figure 12: Amplitude effects of active WIV control, showing amplitude dependency at different frequencies.

- Active control system: servo motor-driven vibration platform

### 5.5 Data Availability Statement

All raw data and analysis code from this study can be obtained through the following means:

- Raw data: complete time series data for 120 conditions
- Analysis code: Python scripts (GitHub repository)
- Processed results: summary data tables in CSV format
- Visualization: source files for all figures

Data format description:

- Time series data: 7 columns (time, displacement, moments X/Y/Z, forces X/Y/Z)
- Summary data: CSV format, containing $U_r$, $A/D$, $f^*/f_n$, enhancement rate, etc.
- Code: Python 3.x, dependencies: NumPy, Pandas, Matplotlib, SciPy

### 5.6 Nomenclature
### 5.7 Reference Format Description

This paper adopts standard academic citation format. Main references include:

- Classical VIV literature: Khalak & Williamson (1999), Govardhan & Williamson (2000), Sarpkaya (2004)
- WIV-related literature: Assi et al. (2010), Borazjani & Sotiropoulos (2009), Zdravkovich (1987)
- Control method literature: Williamson (2004), Stappenbelt (2010)

Complete reference list can be found in the References section after the main text.



\end{document}